\documentclass[graybox]{SNmult}

\usepackage{makeidx}         
\usepackage{graphicx}        
\usepackage{multicol}        
\usepackage[bottom]{footmisc}

\usepackage{newtxtext}       %
\usepackage[varvw]{newtxmath}       

\usepackage{bm}
\usepackage{listings}
\usepackage{chapterbib}
\usepackage{booktabs}
\usepackage{subfigure}
\usepackage{algorithm}
\usepackage{algorithmic}

\usepackage{tcolorbox}
\usepackage{xcolor}
\tcbuselibrary{listings,skins}
\usepackage[numbered,framed,autolinebreaks,useliterate]{matlab-prettifier}

\lstdefinestyle{mystyle}{
numbers=left,
numberstyle=\small,
numbersep=7pt,
language=Matlab,
style=Matlab-editor,
basicstyle=\ttfamily\tiny,
xleftmargin=-12pt,
aboveskip=-6pt,
belowskip=-6pt,
frame=none
}

\makeindex             

\begin{document}

\title*{Characterization of Thermal Systems from Noisy and Low-resolution Measurements Using Dynamic Mode Decomposition}

\titlerunning{Character. Therm. Syst. from Noisy and Low-resolution Measurements Using DMD}

\author{Maria Eduarda Preuss Silva\orcidID{0009-0002-6872-9079},
Lucas Simon Araujo\orcidID{0009-0003-5809-4353},
Fernanda Thais Colombo\orcidID{0000-0002-5988-6067},\\
Americo Cunha Jr\orcidID{0000-0002-8342-0363},
and Samuel da Silva\orcidID{0000-0001-6430-3746}}

\authorrunning{M. E. P. Silva et al.}

\institute{M. E. P. Silva
\at Universidade Estadual Paulista --- UNESP, Ilha Solteira, Brazil 
\and 
L. S. Araujo 
\at Universidade Estadual Paulista --- UNESP, Ilha Solteira, Brazil 
\and 
F. T. Colombo
\at Universidade Estadual Paulista --- UNESP, Ilha Solteira, Brazil 
\and 
A. Cunha~Jr 
\at Laboratório Nacional de Computação Científica --- LNCC, Petrópolis, Brazil\\ \& Universidade do Estado do Rio de Janeiro --- UERJ, Rio de Janeiro, Brazil\\
\email{americo@lncc.br}
\and 
S. da Silva 
\at Universidade Estadual Paulista --- UNESP, Ilha Solteira, Brazil\\ \email{samuel.silva13@unesp.br}
}

\maketitle

\label{Chapter:09}

\abstract*{Thermal monitoring in practical applications is often constrained by sparse sensing, measurement noise, and limited spatial resolution, which hinder the identification of heat transfer dynamics. In such settings, calibrating high-fidelity physical models is computationally demanding, motivating data-driven approaches. Dynamic Mode Decomposition (DMD) provides a framework for extracting spatiotemporal structures from measurement data, but its standard formulation is sensitive to noise and degraded observations. This chapter examines the use of DMD under these constraints, focusing on preprocessing and truncation strategies that affect stability and interpretability. Two cases are considered: forced convection with thermocouple data and transient heat conduction from degraded thermal images. The number of retained modes is treated as a modeling parameter that governs the trade-off between reconstruction fidelity and noise sensitivity. The results indicate that DMD recovers dominant thermal behavior from both sparse and degraded datasets when the truncation level is appropriately selected. Low-rank models provide stable but simplified descriptions, while higher-rank models improve spatial detail at the cost of increased noise sensitivity.}

\abstract{Thermal monitoring in practical applications is often constrained by sparse sensing, measurement noise, and limited spatial resolution, which hinder the identification of heat transfer dynamics. In such settings, calibrating high-fidelity physical models is computationally demanding, motivating data-driven approaches. Dynamic Mode Decomposition (DMD) provides a framework for extracting spatiotemporal structures from measurement data, but its standard formulation is sensitive to noise and degraded observations. This chapter examines the use of DMD under these constraints, focusing on preprocessing and truncation strategies that affect stability and interpretability. Two cases are considered: forced convection with thermocouple data and transient heat conduction from degraded thermal images. The number of retained modes is treated as a modeling parameter that governs the trade-off between reconstruction fidelity and noise sensitivity. The results indicate that DMD recovers dominant thermal behavior from both sparse and degraded datasets when the truncation level is appropriately selected. Low-rank models provide stable but simplified descriptions, while higher-rank models improve spatial detail at the cost of increased noise sensitivity.}

\section{Introduction}
\label{chap:intro}

Thermal monitoring and analysis in industrial systems are often constrained by limited sensing, measurement noise, and reduced spatial resolution. These conditions are common in applications such as thermal energy conversion, refrigeration, and chemical processing, where reliable diagnostics and efficient operation depend on incomplete and imperfect data \cite{Sherif2022, Jonsson2007}. Under such constraints, constructing accurate models from first principles becomes challenging, motivating the use of data-driven approaches that can extract dynamical information directly from measurements \cite{Brunton2019}.

Dynamic Mode Decomposition (DMD), introduced by \cite{Schmid2010}, provides a framework for identifying spatiotemporal structures from time-resolved data. Rooted in Koopman operator theory \cite{Brunton2019,DMDbook}, DMD approximates the evolution of complex systems through linear representations derived from snapshot sequences. Its ability to reveal coherent structures and associated temporal behavior has led to applications in diverse areas, including epidemiology \cite{Takahashi2021}, structural dynamics \cite{Saito2020}, and heat transfer \cite{SilvaCIC2024, SilvaERMAC2024}. However, when applied to thermal systems with noisy measurements and limited resolution, standard DMD formulations may produce modes that are sensitive to noise or difficult to interpret physically.

This chapter examines how DMD can be adapted and interpreted under such practical measurement constraints. Rather than introducing a fundamentally new algorithm, the focus is on a structured application of DMD that combines preprocessing, rank selection, and mode interpretation to improve reliability in scenarios with sparse sensors and degraded thermal images \cite{Baddoo2023, Brunton2019}. The objective is to assess how these methodological choices influence the extraction of dominant thermal dynamics and the reconstruction of temperature fields from imperfect data.

Two representative thermal systems are used to support this analysis. The first involves forced convection, with temperature measurements obtained from a sensor-instrumented test bench, yielding a low-dimensional, noisy representation of the thermal field. The second considers transient heat conduction in a metallic plate, using thermal images intentionally degraded by noise addition and resolution reduction. These cases allow the evaluation of DMD under distinct measurement modalities, highlighting how data quality and dimensionality affect reconstruction and mode selection.

The results indicate that DMD can recover coherent spatiotemporal structures from both sparse sensor data and degraded images, provided that the truncation level is carefully selected. Low-rank approximations capture the dominant behavior but may oversimplify spatial variations, whereas higher-rank models improve detail at the expense of increased sensitivity to noise. These observations emphasize the role of rank selection as a modeling parameter rather than a purely numerical choice.

Within this scope, the chapter positions DMD as a practical tool for analyzing thermal systems when detailed physical models are unavailable or difficult to calibrate. By operating directly on measured data, the approach provides insight into the interaction of variables such as temperature, flow, and heat transfer without requiring explicit governing equations, while remaining sensitive to the limitations imposed by measurement quality.

The chapter is organized as follows. Section~\ref{sec:2} formulates the problem and outlines the guiding hypotheses. Section~\ref{sec:3} presents the DMD formulation adopted and discusses its application to noisy and low-resolution data. Section~\ref{sec:4} reports the results for forced convection and transient heat conduction. Section~\ref{sec:5} summarizes the main findings and discusses their implications.

\section{Problem Statement and Hypotheses}
\label{sec:2}

This section formulates the problem addressed in this chapter, focusing on the identification of thermal dynamics from incomplete, noisy measurements. In practical industrial settings, temperature information is typically obtained from a limited number of sensors or from imaging devices with restricted spatial resolution. These measurements are further affected by noise, which obscures the underlying spatiotemporal structures governing heat transfer processes. As a result, reconstructing and interpreting the system dynamics from available data becomes a challenging task, both because the observable information is incomplete and because noise contaminates the structures of interest.

Let $\bm{y}(t) \in \mathbb{R}^m$ denote the measured observable at time $t$, representing either a collection of sensor readings or a vectorized thermal image of dimension $m$. In practice, $\bm{y}(t)$ is related to the true thermal state $\bm{x}(t) \in \mathbb{R}^n$, with $m \leq n$, through
\begin{equation}
    \bm{y}(t) = \bm{C}\bm{x}(t) + \bm{\eta}(t),
\end{equation}
where $\bm{C} \in \mathbb{R}^{m \times n}$ is an observation operator that maps the full state to the available measurements, and $\bm{\eta}(t) \in \mathbb{R}^m$ represents additive measurement noise. The operator $\bm{C}$ encodes both sparse sensor placement (rows selecting specific spatial locations) and resolution reduction (spatial averaging over image pixels). This formulation makes explicit the two distinct sources of information loss: dimensionality reduction through $\bm{C}$ and noise contamination through $\bm{\eta}$.

The objective is to construct a reduced-order representation of the temporal evolution directly from the observables $\bm{y}(t)$, without access to the full state $\bm{x}(t)$. Specifically, the goal is to approximate the dynamics through a linear mapping
\begin{equation}
    \bm{y}_{k+1} \approx \bm{A} \bm{y}_k,
\end{equation}
where $\bm{A}$ is an operator identified from time-resolved snapshot data, and the discrete index $k$ is related to time by $t_k = k \Delta t$, with $\Delta t$ denoting the sampling interval. This representation enables the extraction of dynamic modes and their temporal behavior, providing a basis for reconstructing and interpreting the evolution of the thermal field from imperfect observations alone.

DMD is employed to identify $\bm{A}$ from snapshot data. When applied to measurements affected by noise and limited resolution, however, standard DMD formulations may produce modes that are sensitive to measurement perturbations or lack clear physical interpretation. These issues are closely linked to the choice of truncation rank and to data preprocessing, both of which influence the balance between capturing relevant dynamics and amplifying noise. The analysis is guided by the following hypotheses, each of which is evaluated against the experimental results in Section~\ref{sec:5}:
\begin{itemize}
    \item \textbf{H1:} The dominant thermal dynamics can be represented by a low-dimensional set of coherent modes, even when measurements are noisy and spatially limited. This is assessed by examining whether a small number of DMD modes accounts for the dominant fraction of variance in the reconstructed fields.

    \item \textbf{H2:} The truncation rank governs a trade-off between reconstruction fidelity and sensitivity to noise. This is evaluated by comparing normalized reconstruction errors across truncation levels and examining whether higher-rank models exhibit increased noise amplification.
\end{itemize}

These hypotheses are examined through two case studies involving forced convection and transient heat conduction, using both sparse sensor data and degraded thermal images. The analysis focuses on how measurement limitations and modeling choices affect DMD's ability to produce consistent reduced-order representations of thermal dynamics.

\section{DMD for Noisy and Low-Resolution Thermal Data}
\label{sec:3}

This section outlines the DMD formulation adopted in this chapter, with emphasis on its application to thermal data affected by noise and limited spatial resolution. Rather than revisiting the theoretical foundations of DMD (see Chapter~\ref{Chapter:08}), the focus here is on the practical steps that influence the stability and interpretability of the decomposition under realistic measurement conditions.

Let $\bm{y}_k \in \mathbb{R}^m$ denote the measured thermal observable at 
discrete time $t_k = k\Delta t$. For sensor-based measurements, $\bm{y}_k$ 
collects the readings from $m$ thermocouples at time step $k$. For thermal 
images, $\bm{y}_k$ is obtained by vectorizing the $p \times q$ pixel array, 
giving $m = pq$.

From the preprocessed sequence of $N$ snapshots, two matrices are 
constructed:
\begin{equation}
    \mathbf{Y} =
    \begin{bmatrix}
        \bm{y}_1 & \bm{y}_2 & \cdots & \bm{y}_{N-1}
    \end{bmatrix}, \quad
    \mathbf{Y}' =
    \begin{bmatrix}
        \bm{y}_2 & \bm{y}_3 & \cdots & \bm{y}_N
    \end{bmatrix},
\end{equation}
so that $\mathbf{Y}' \approx \mathbf{A}\mathbf{Y}$ defines the 
linear regression problem.

A direct least-squares estimate of $\mathbf{A}$ may become sensitive to noise and limited sampling, leading to overfitting and the identification of modes that do not reflect coherent thermal structures. To mitigate these effects, a reduced-order representation is constructed using the truncated Singular Value Decomposition (SVD) of $\mathbf{Y}$,
\begin{equation}
\mathbf{Y} \approx \tilde{\mathbf{U}} \tilde{\mathbf{\Sigma}} \tilde{\mathbf{V}}^T.
\end{equation}

Here, $\tilde{\mathbf{U}} \in \mathbb{R}^{m \times r}$, 
$\tilde{\mathbf{\Sigma}} \in \mathbb{R}^{r \times r}$, and 
$\tilde{\mathbf{V}} \in \mathbb{R}^{(N-1) \times r}$ denote the truncated SVD components obtained by retaining the $r$ dominant singular values and corresponding singular vectors. The reduced operator is then defined as
\begin{equation}
\tilde{\mathbf{A}} = \tilde{\mathbf{U}}^T \mathbf{Y}' \tilde{\mathbf{V}} \tilde{\mathbf{\Sigma}}^{-1}.
\end{equation}

The eigen-decomposition of $\tilde{\mathbf{A}}$ yields the DMD eigenvalues and modes, which encode the temporal evolution and spatial structures present in the data. The thermal field can then be reconstructed as a superposition of these modes, providing a compact description of the underlying dynamics.

Within this framework, the truncation rank $r$ becomes a key modeling parameter. Low values of $r$ filter measurement noise and produce stable representations, but may suppress spatial variability and transient features. In contrast, higher values of $r$ increase the level of detail in the reconstruction but also introduce sensitivity to noise and potential spurious oscillations. The selection of $r$ therefore reflects a trade-off between reconstruction fidelity and robustness, and is treated here as an integral part of the modeling process rather than a purely numerical choice.

In addition to rank selection, preprocessing is central to ensuring consistent input data. For sensor-based measurements, preprocessing includes temporal alignment and normalization to reduce inconsistencies across channels. For thermal images, this involves reshaping the data into a vector, adjusting the spatial resolution, and accounting for noise contamination. These steps do not alter the underlying dynamics but affect how reliably they can be extracted by the decomposition.

Taken together, these elements define a practical workflow for applying DMD to thermal systems under measurement constraints. The combination of dimensionality reduction, controlled rank selection, and data preprocessing enables the extraction of coherent spatiotemporal patterns from datasets that are sparse, noisy, or degraded. The impact of these choices is examined in the next section through experimental studies of forced convection and transient heat conduction.

\section{Results and Discussion}
\label{sec:4}

This section evaluates DMD's behavior under two distinct measurement scenarios. In the first, spatial resolution is severely limited by the number of available sensors; in the second, both noise and spatial resolution degradation affect a high-dimensional image dataset. In both cases, the analysis focuses on how the truncation rank influences the balance between capturing thermal structures and amplifying measurement noise, treating $r$ as a modeling parameter rather than a purely numerical choice.

\subsection{Forced convection under noisy temperature measurements}
\label{sec:41}

\begin{figure}[h!]
\sidecaption
\includegraphics[width=1\textwidth]{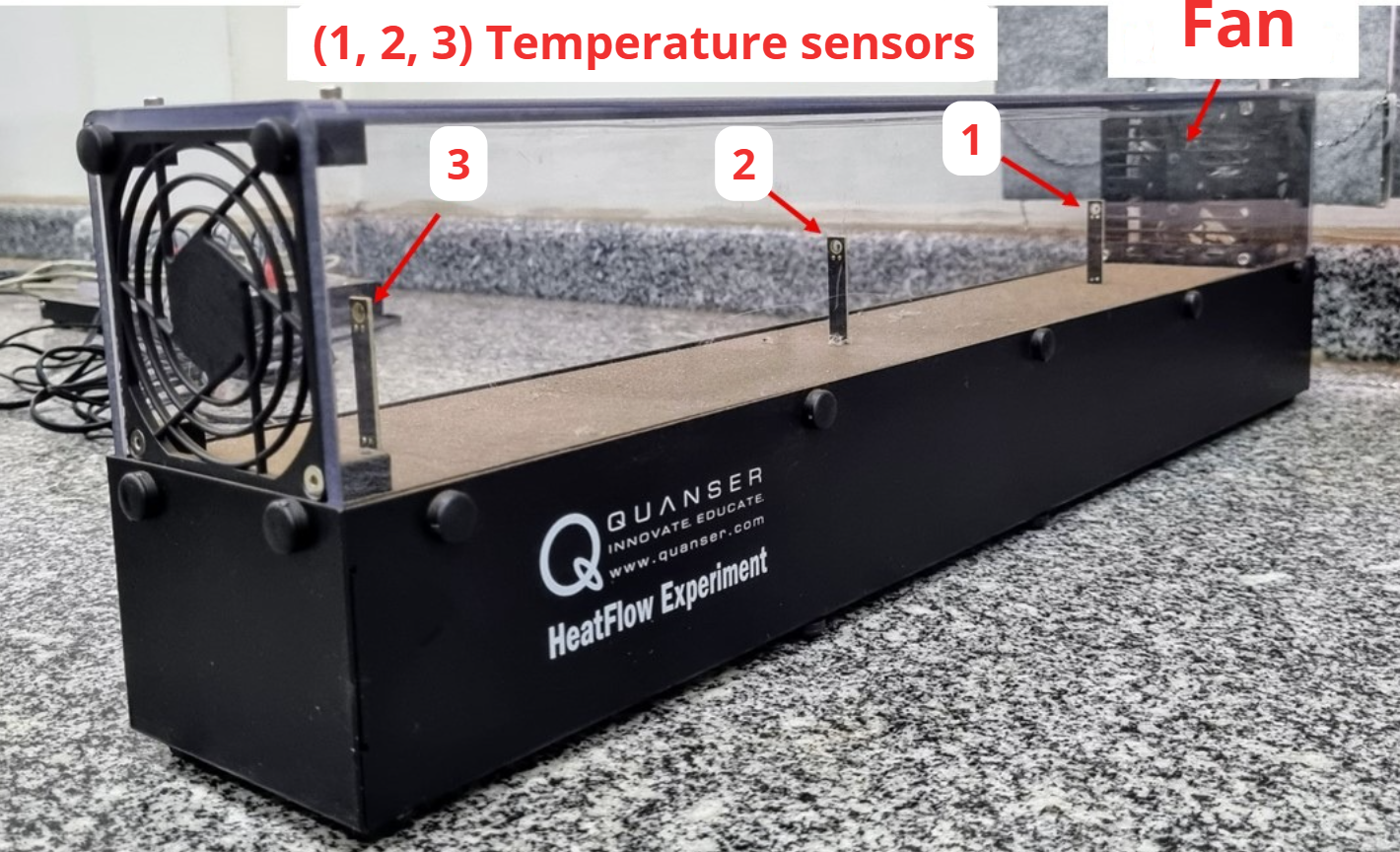}
\Description{Experimental setup used for forced convection measurements.}
\caption{Forced convection experimental bench (Quanser HeatFlow). Three thermocouple sensors (T1, T2, T3) are positioned along the test section from the fan end toward the heater, providing a spatially sparse representation of the axial temperature distribution.}
\label{fig:bancada}
\end{figure}

The experimental setup consists of a Quanser forced convection bench (Fig.~\ref{fig:bancada}) equipped with three thermocouples distributed along the test section. A resistive heater establishes a thermal gradient along the bench, while a fan controlled by an analog amplifier regulates the airflow. Fan speed was monitored with a tachometer to ensure consistent operating conditions. Temperature signals were sampled at 500 Hz, providing sufficient temporal resolution to capture the system's dominant thermal dynamics.

With only three sensors available, the snapshot matrix has dimension $m = 3$, and the maximum number of extractable DMD modes is three. Reconstructions are therefore performed for $r = 1$, $r = 2$, and $r = 3$ to span the full range of representational capacity available under this spatial constraint. In each case, the left panel of the corresponding figure shows the reconstructed temperature field as a function of sensor position and time, and the right panel shows the spatial profile of the extracted dynamic mode or modes.

\begin{figure}[h!]
\sidecaption
\includegraphics[width=1\textwidth]{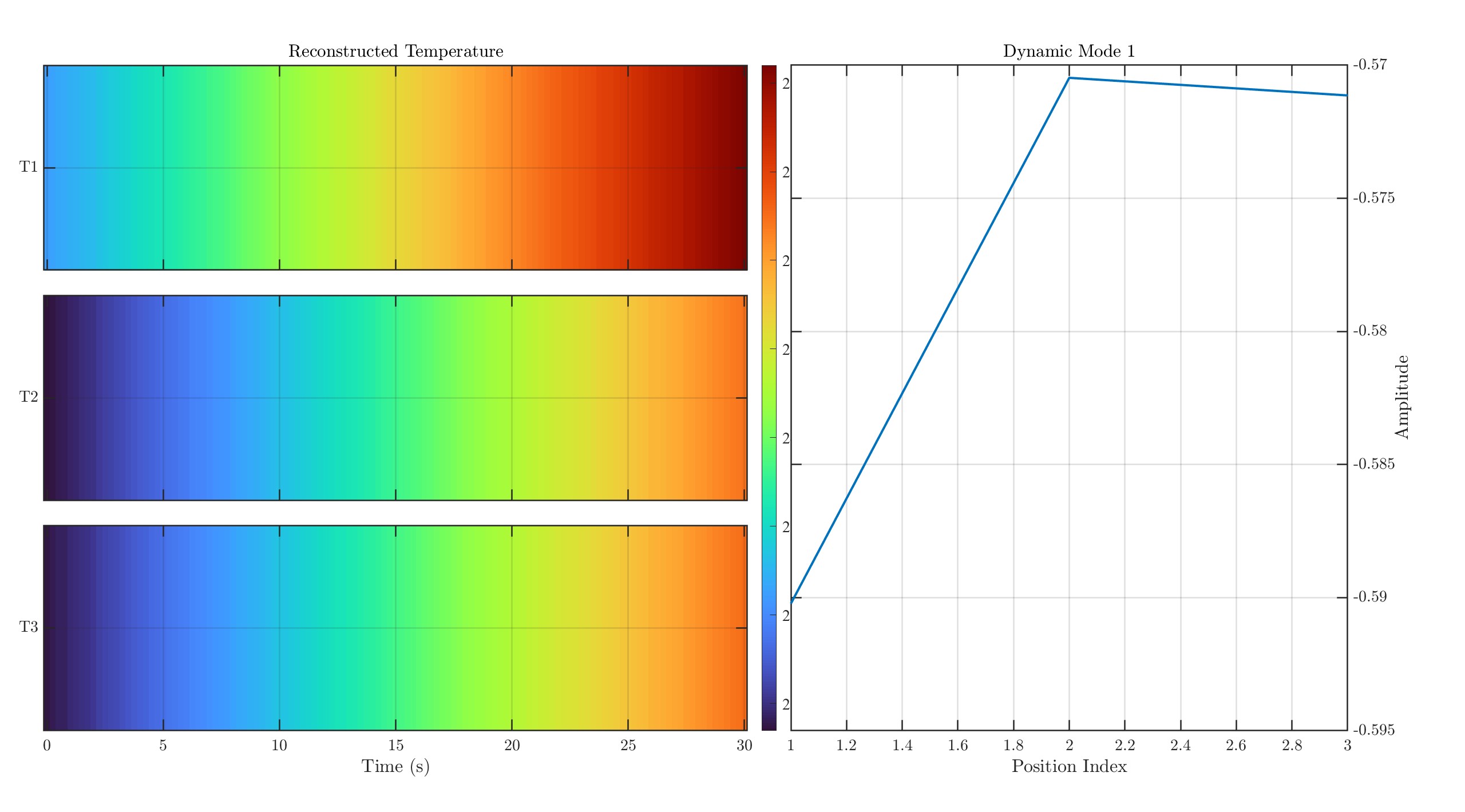}
\Description{Temperature field reconstruction and dynamic mode for r=1.}
\caption{Reconstruction using $r = 1$ mode. The spatial profile of the mode reflects the mean axial gradient, but the uniform temporal behavior across sensors fails to represent the differential convective response between upstream and downstream positions.}
\label{fig:rec1}
\end{figure}

As depicted in Fig.~\ref{fig:rec1}, for $r = 1$, the single extracted mode has a monotonically increasing spatial profile from sensor T1 to T3, reflecting the steady axial temperature gradient established by the heater. The reconstructed field evolves uniformly in time across all three sensors, capturing the global heating trend but producing identical temporal behavior at every spatial location. This representation fails to resolve the differential response of each sensor: in particular, T1 (closest to the fan) responds more rapidly to airflow changes than T3 (closest to the heater), a feature that a single-mode approximation cannot encode. The result is a spatially averaged description of the thermal evolution that suppresses the convective transport signature.

\begin{figure}[h!]
\sidecaption
\includegraphics[width=1\textwidth]{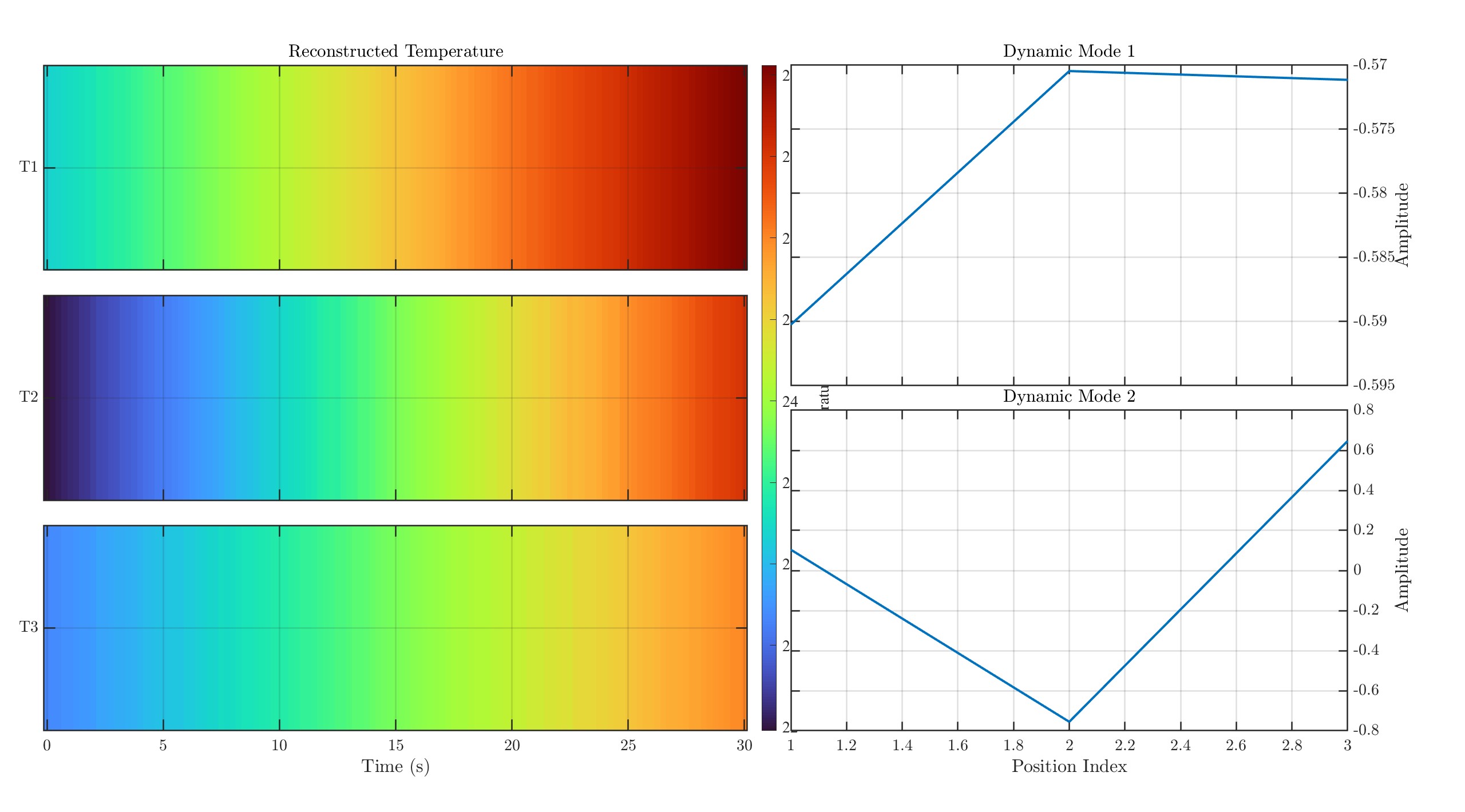}
\Description{Temperature field reconstruction and dynamic mode for r=2.}
\caption{Reconstruction using $r = 2$ modes. The second mode captures the differential axial response between sensors, encoding the spatial signature of convective heat transport that is absent in the single-mode approximation.}
\label{fig:rec2}
\end{figure}

For $r = 2$ (Fig.~\ref{fig:rec2}), the second mode introduces a spatial profile that changes sign between sensors, consistent with the differential temperature response driven by convective transport. The reconstructed field now exhibits distinct temporal trajectories at each sensor location: T1 and T3 diverge during transient phases, capturing the lag between the upstream cooling effect of the fan and the downstream influence of the heater. This two-mode representation recovers the qualitative structure of forced convection along the bench and reduces the visible discrepancy between the reconstruction and the measured signals, particularly during transient periods when airflow changes produce spatially non-uniform responses.

\begin{figure}[t]
\sidecaption
\includegraphics[width=1\textwidth]{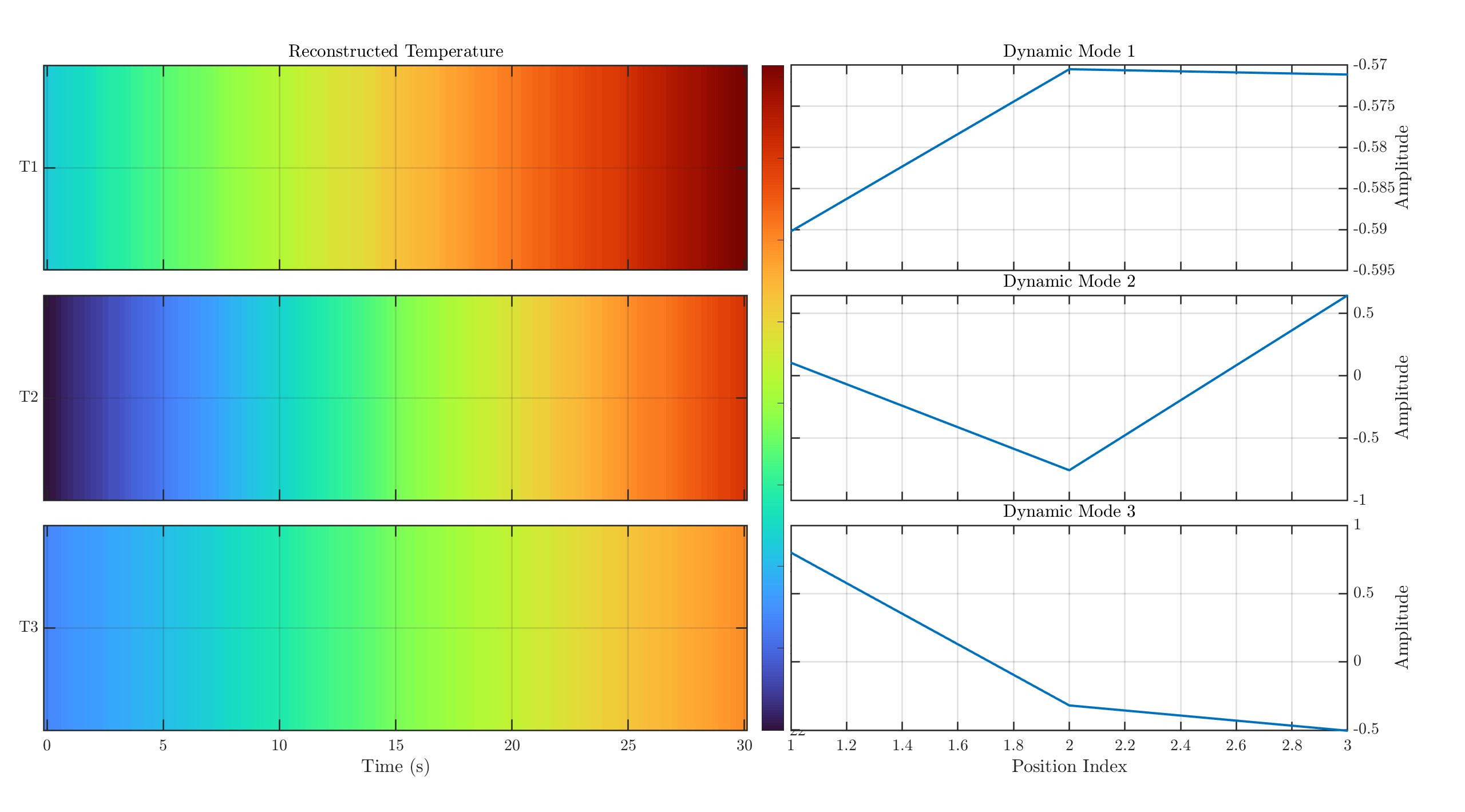}
\Description{Temperature field reconstruction and dynamic mode for r=3.}
\caption{Reconstruction using $r = 3$ modes. The full modal basis reproduces the measured signals at all three sensors; however, since $r = 3$ exhausts the available spatial dimension, the third mode cannot be distinguished from noise through truncation alone and requires physical interpretation to assess its relevance.}
\label{fig:rec3}
\end{figure}

When $r = 3$ (Fig.~\ref{fig:rec3}), all three available modes are retained, and the reconstruction closely follows the measured signals at all sensor locations. The third mode captures localized, lower-amplitude variations that the two-mode model leaves as residual. Comparing the spatial profiles of the three modes reveals a clear hierarchy: the first mode represents the dominant global gradient, the second encodes the primary convective asymmetry, and the third reflects finer variations that, at the noise levels present in this dataset, are difficult to distinguish unambiguously from measurement fluctuations. Because $r = 3$ exhausts the available spatial degrees of freedom, no truncation filtering is applied, and the decomposition fits the data exactly, including any noise contribution. This means the improvement in reconstruction fidelity at $r = 3$ relative to $r = 2$ should be interpreted cautiously: part of the gain reflects genuine thermal dynamics, while the rest reflects overfitting to noise in the third sensor channel.

Taken together, these results illustrate a fundamental constraint of DMD under extreme spatial sparsity: when the number of sensors equals the maximum model order, rank truncation ceases to serve as a noise filter. In this regime, the value of DMD lies not in denoising but in decomposing the available measurements into interpretable contributions. The mode shapes obtained here are physically consistent with the expected structure of forced convection — a dominant gradient mode, a convective asymmetry mode, and a residual mode — supporting the use of DMD as a diagnostic tool even when spatial coverage is minimal.

\subsection{Transient heat conduction from low-resolution thermal images}
\label{sec:42}

\begin{figure}[t]
\sidecaption
\includegraphics[width=0.65\textwidth]{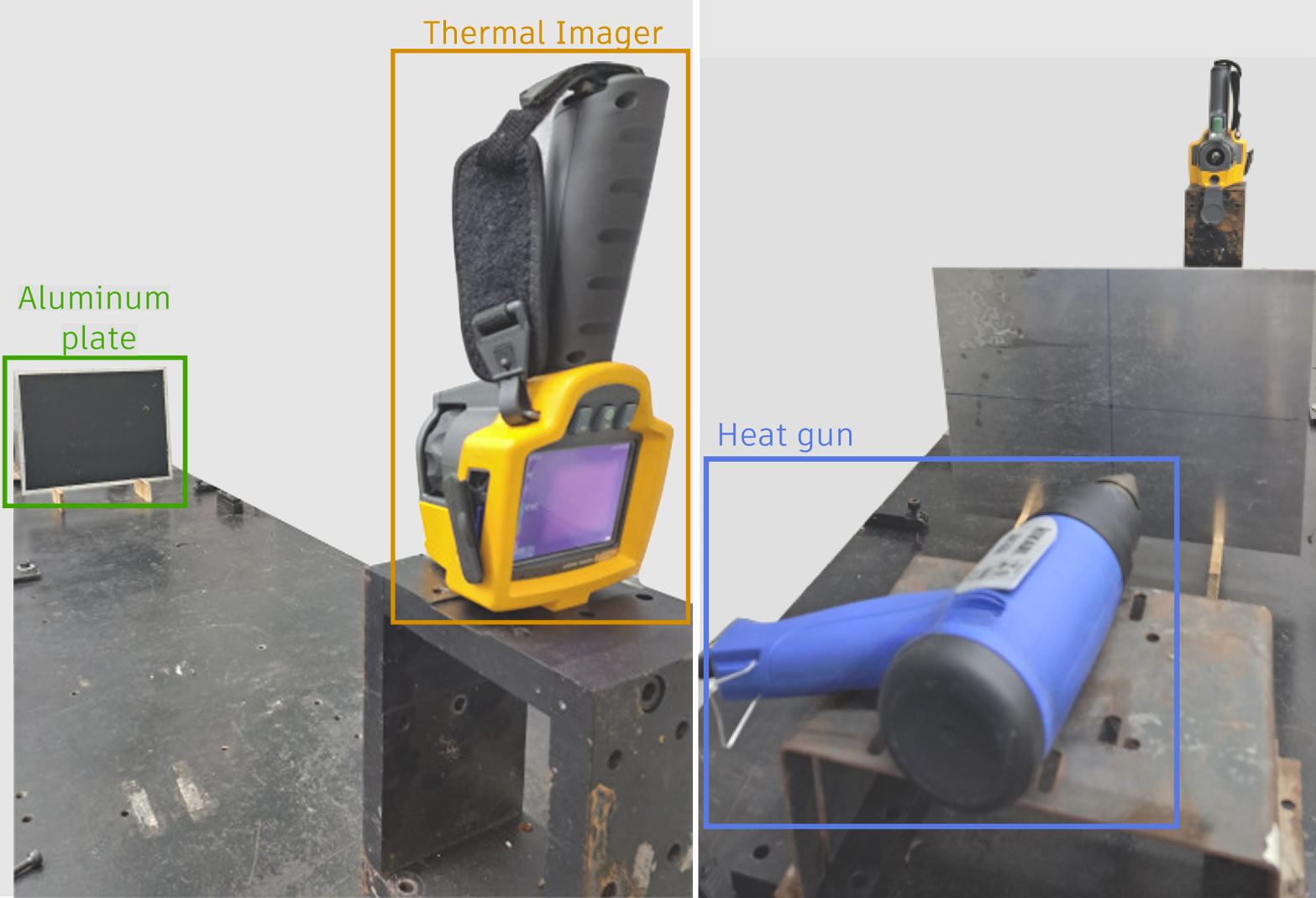}
\Description{Experimental setup for thermal imaging.}
\caption{Experimental setup for transient heat conduction. Front and rear views show the aluminum plate, the Fluke Ti25 thermal imager, and the heat gun used to locally excite the plate at its center.}
\label{fig:esquema1}
\end{figure}

The second case study involves transient heat conduction in a flat aluminum plate (305~mm $\times$ 250~mm $\times$ 1.5~mm), whose surface was coated with matte black paint to increase emissivity and reduce reflectivity for infrared measurements. Thermal excitation was provided by a Hikari HK-508 heat gun (nominal power 1500~W) applied to the central region of the plate for 150~s, followed by a 300~s natural cooling phase under ambient conditions. The temperature field was recorded using a Fluke Ti25 thermal imager at 10~s intervals, yielding a sequence of 45 frames that capture both the heating and cooling transients (Fig.~\ref{fig:esquema1}).

To emulate practical measurement degradation, the original thermal images were processed in two ways before DMD analysis: Gaussian noise with variance $\sigma^2 = 0.005$ was added to the normalized pixel values, and spatial resolution was reduced by a factor of 0.9 through resampling. Each processed image was then vectorized and assembled into the snapshot matrices $\mathbf{Y}_1$ and $\mathbf{Y}_2$ following the procedure described in Section~\ref{sec:3}. Reconstructions are evaluated for $r = 1$, $r = 10$, and $r = 40$.

\begin{figure}[h!]
\sidecaption
\includegraphics[width=1\textwidth]{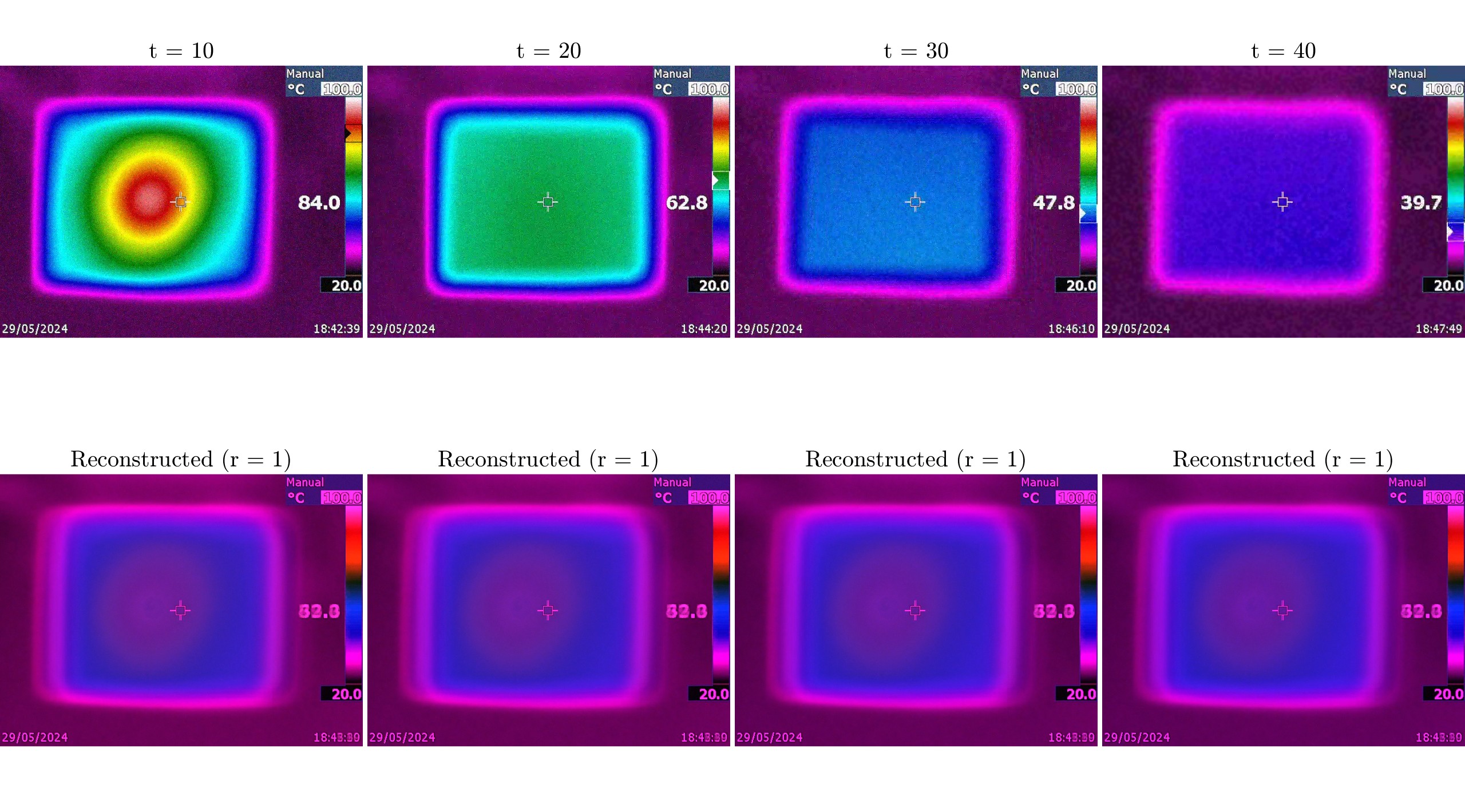}
\Description{Mode and reconstruction using one mode.}
\caption{Dynamic mode (top) and image reconstruction (bottom) for $r = 1$. The mode captures the dominant spatial envelope of the heated region; the reconstruction preserves the global temperature distribution but does not resolve the temporal progression of the diffusion front.}
\label{fig:frame1}
\end{figure}

For $r = 1$ (Fig.\ref{fig:frame1}), the dominant mode exhibits a broad spatial envelope centered on the heated region, with amplitude decaying toward the plate edges. This structure corresponds to the lowest spatial frequency of the thermal field and reflects the diffusive spread of heat from the source. The reconstructed images (Fig.~\ref{fig:frame1} bottom) show that this single-mode approximation preserves the overall shape of the heated zone at each time step but assigns nearly identical spatial distributions at $t = 10$, 20, 30, and 40~s, failing to represent the temporal evolution of the diffusion front as it progresses toward the plate boundary. The cooling phase is similarly oversimplified: the reconstruction captures the decreasing temperature amplitude but cannot represent the spatial redistribution of heat that occurs as conduction and convection interact during natural cooling.

\begin{figure}[t]
\sidecaption
\includegraphics[width=1\textwidth]{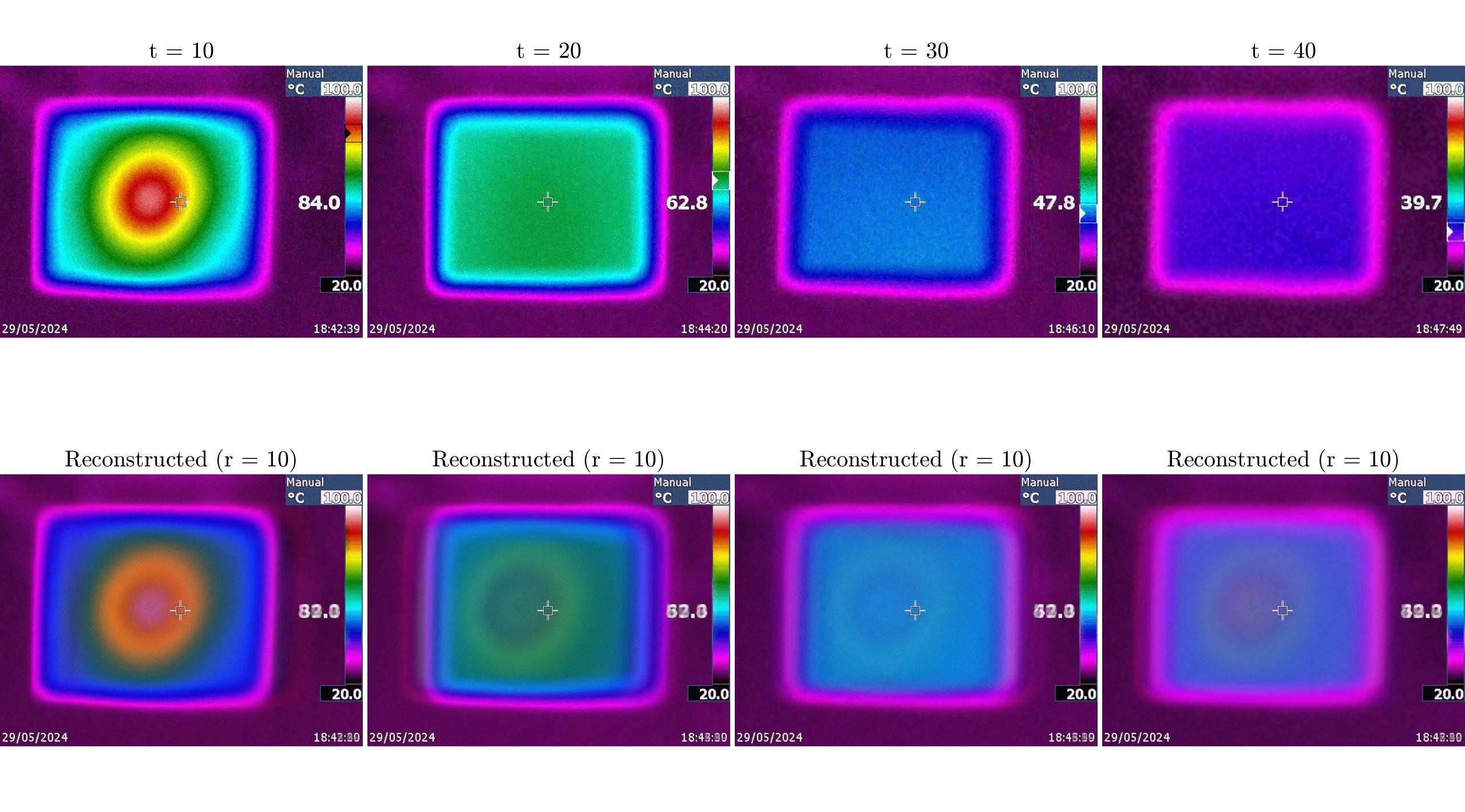}
\Description{Mode and reconstruction using ten modes.}
\caption{Dynamic modes (top) and image reconstruction (bottom) for $r = 10$. The multimodal representation recovers the transient diffusion front and the spatial gradient between the heated zone and the plate boundaries, features that are absent in the single-mode approximation.}
\label{fig:frame2}
\end{figure}

Taking $r = 10$ (Fig.~\ref{fig:frame2}), the ten extracted modes span a range of spatial frequencies. The lower-indexed modes retain smooth, large-scale structures similar to the dominant mode, while higher-indexed modes exhibit increasingly fine spatial oscillations. The reconstructed images now show clear differences across time steps: the localized heating peak visible at $t = 10$~s broadens and weakens progressively, and the spatial gradient between the central heated zone and the cooler plate boundaries is reproduced with substantially greater fidelity than in the single-mode case. Comparing the reconstructed and original images, the main spatiotemporal features of the conduction and cooling process are recovered, including the asymmetric spread of the thermal front and the preferential cooling along the plate edges. The mode profiles do not exhibit the irregular oscillatory features associated with noise amplification, suggesting that the ten-mode truncation retains predominantly physical content at this noise level.

\begin{figure}[t]
\sidecaption
\includegraphics[width=1\textwidth]{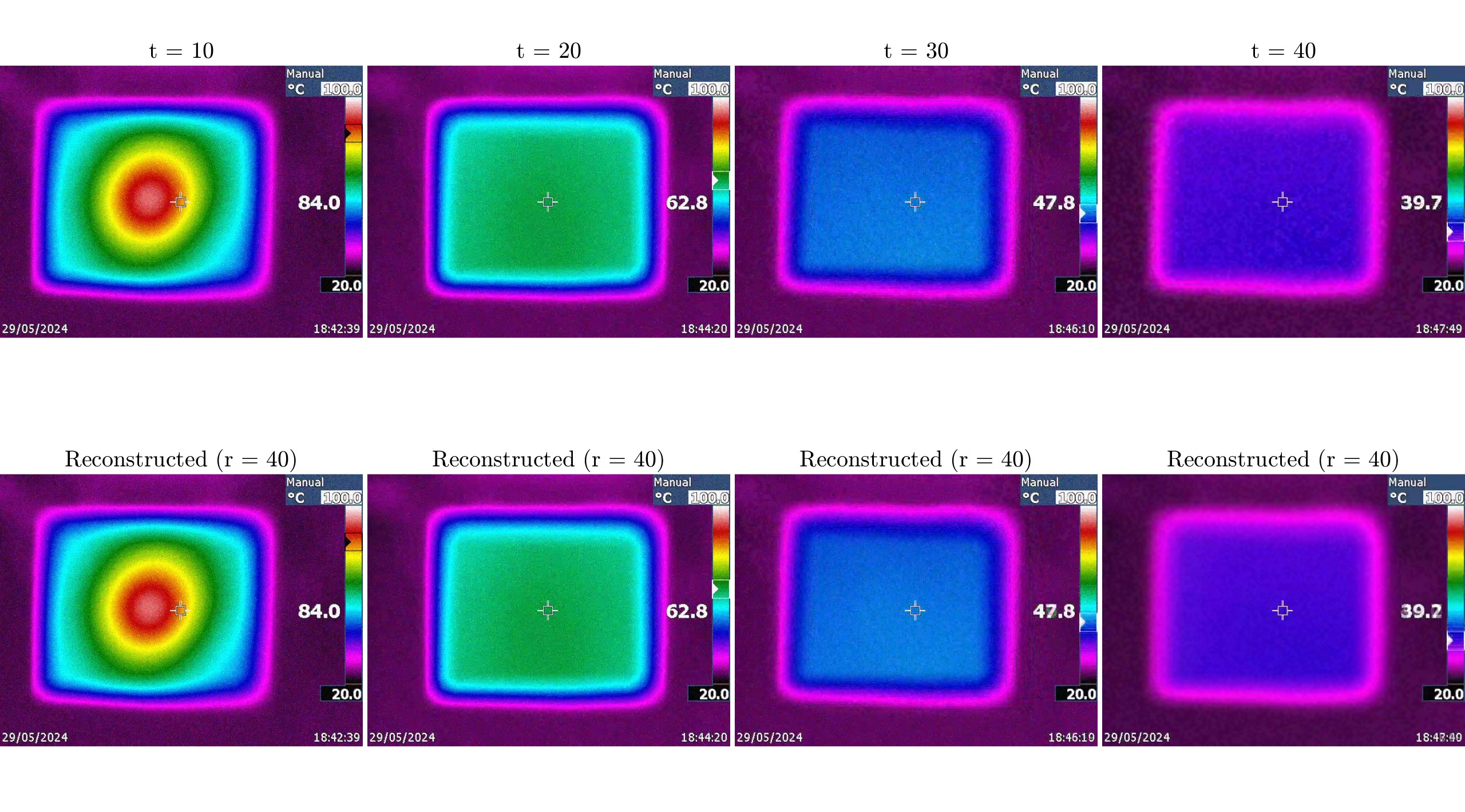}
\Description{Mode and reconstruction using forty modes.}
\caption{Dynamic modes (top) and image reconstruction (bottom) for $r = 40$. The reconstructed field closely resembles the original images, but higher-indexed mode profiles exhibit irregular oscillations inconsistent with the expected spatial structure of heat conduction, indicating that the decomposition at this rank incorporates noise-amplified components.}
\label{fig:frame3}
\end{figure}

Choosing $r = 40$ (Fig.~\ref{fig:frame3}), the reconstructed images closely resemble the original degraded frames, with sharper spatial gradients and finer detail near the plate edges. However, examining the mode profiles at higher indices reveals a qualitative change: while the lower modes retain spatially coherent structures, modes beyond approximately index 20 exhibit irregular, high-frequency spatial oscillations that do not correspond to physically expected heat conduction patterns. These features are consistent with the amplification of the added Gaussian noise, which distributes energy across many singular vectors and becomes progressively more prominent as $r$ increases beyond the noise floor of the singular value spectrum. The visual improvement in the reconstructed images at $r = 40$ relative to $r = 10$ is therefore partially attributable to fitting noise rather than to the recovery of additional physical dynamics.

The progression from $r = 1$ to $r = 10$ to $r = 40$ illustrates the central modeling tension in DMD applied to degraded data. The improvement from one to ten modes is qualitatively substantial: the additional modes encode physically meaningful spatial structures that the single-mode representation cannot capture, and the reconstructed dynamics are consistent with the expected behavior of transient heat conduction. The improvement from $r = 10$ to $r = 40$ is more ambiguous: the reconstructed images gain visual fidelity, but the mode shapes suggest that part of this gain comes from fitting measurement noise. This distinction is important in practice because noise-fitted modes do not generalize: a model that includes them will fail to predict the thermal evolution at time steps not used to construct the snapshot matrices.

Across both case studies, the results are consistent with the framework established in Section~\ref{sec:2}.  These observations are consistent with Hypothesis H1, as the dominant thermal dynamics are captured by a small number of coherent modes.  This behavior supports Hypothesis H2, highlighting the trade-off between reconstruction fidelity and noise sensitivity as the truncation rank increases.

In both the forced convection and heat conduction cases, the first ten modes capture the primary spatial and temporal structures, while higher truncation levels introduce components that are increasingly influenced by measurement noise.

\section{Final Remarks}
\label{sec:5}

This chapter examined the use of Dynamic Mode Decomposition as a data-driven approach for analyzing and reconstructing thermal systems under measurement constraints. The focus was on conditions commonly encountered in practice, including noisy data, sparse sensing, and limited spatial resolution, where the reliability of model-based analysis is inherently reduced.

The results from the forced convection case indicate that DMD can recover the system's dominant dynamical behavior even with only a few temperature measurements. Low-order representations capture the global thermal trend in a stable manner, while higher-order models introduce spatial variability and improve fidelity to the measured signals. This progression highlights the role of truncation rank as a mechanism for controlling the level of detail in the reconstruction, with increasing model order also reducing the implicit filtering of noise.

A similar pattern is observed in the transient heat conduction case. When applied to degraded thermal images, DMD reconstructs the main spatiotemporal structures of the temperature field despite noise contamination and reduced resolution. Low-rank approximations provide smooth representations of the dominant heat distribution, whereas higher-rank models recover finer spatial features at the cost of increased sensitivity to noise. Intermediate truncation levels offer a consistent balance, preserving relevant thermal patterns while limiting the influence of measurement perturbations.

Taken together, these results indicate that DMD's effectiveness in thermal applications depends not only on available data but also on the decomposition configuration. In particular, the truncation rank functions as a modeling parameter that governs the trade-off between reconstruction fidelity and robustness. When selected appropriately, DMD extracts coherent thermal structures from both sparse sensor measurements and degraded image data, without requiring explicit knowledge of the governing equations.

Within the scope of this study, DMD can be interpreted as a practical tool for reduced-order modeling of thermo-fluid systems when detailed physical models are unavailable or difficult to calibrate. The combination of dimensionality reduction, modal decomposition, and preprocessing provides a structured approach to analyzing thermal dynamics directly from measurements while remaining sensitive to limitations imposed by data quality.

The analysis presented here is limited to two representative experimental configurations and to standard DMD with controlled truncation. Further developments may include the use of advanced DMD variants, systematic strategies for rank selection, and the incorporation of physics-informed constraints. Extensions based on noise-robust variants of DMD may further improve performance under measurement uncertainty. These directions may increase the methodology's robustness and extend its applicability to more complex systems operating under uncertain, data-limited conditions.

\ethics{Code Availability}{The MATLAB code employed in this work is available in the \emph{Dynamic Mode Decomposition Engine (DynaMoDE)} repository at \url{https://github.com/americocunhajr/DynaMoDE}.}

\begin{acknowledgement}
This research was supported by the National Council for Scientific and Technological Development (CNPq, grants 309467/2023-3 and 305476/2022-0); the Coordination for the Improvement of Higher Education Personnel (CAPES, grant 88887.801227/2023-00); Foundation for Research Support of the State of São Paulo (FAPESP, grant 22/16271-2); and the Foundation for Research Support of the State of Rio de Janeiro (FAPERJ, grant 204.477/2024). Additional support was provided by the National Institute of Science and Technology for Smart Structures in Engineering (INCT-EIE), funded by CNPq under grant 406148/2022-8, and by the Minas Gerais Research Support Foundation (FAPEMIG).
\end{acknowledgement}

\ethics{AI-assisted Content Development}{The authors used artificial intelligence tools, including ChatGPT and Grammarly, to support language refinement and improve readability. The scientific content, interpretations, and final text were fully reviewed and validated by the authors, who assume complete responsibility for the manuscript.}

\ethics{Competing Interests}{The authors have no conflicts of interest to declare that are relevant to the content of this chapter.}

\ethics{Copyright notice}{This manuscript has been accepted for publication as a chapter in the book \emph{Scientific Machine Learning for Predictive Modeling: Bridging Data-Driven and Physics-Based Approaches in Computational Science and Engineering}, edited by A. Cunha Jr, F. P. Santos, F. A. Rochinha, A. L. G. A. Coutinho, to be published by Springer Nature. The final authenticated version will be available through Springer Nature.}

\bibliographystyle{spmpsci}
\bibliography{Refs-Chap09}

\end{document}